\documentstyle[aps,epsf]{revtex}

\begin{document}

\title{Integrability and Quantum Phase Transitions in Interacting Boson Models}

\author{J. Dukelsky$^1$, J. M. Arias$^2$, J. E. Garcia-Ramos$^3$ and S. Pittel$^4$}

\address{$^1$ Instituto de Estructura de la Materia, CSIC, Serrano
 $123$, Madrid, Spain \\
 $^2$ Departamento de F\'{\i}sica At\'omica, Molecular y Nuclear, Universidad de Sevilla, Spain\\
 $^3$ Deparatmento de F\'{\i}sica Aplicada, Universidad de Huelva, Spain\\
 $^4$ Bartol Research Institute, University of Delaware, Newark, Delaware 19716, USA}

\maketitle

\begin{abstract}
 The exact solution of the boson pairing hamiltonian
given by Richardson in the sixties is used to study the phenomena of level crossings and quantum
phase transitions in the integrable regions of the $sd$ and $sdg$ interacting boson models.

\end{abstract}

\vspace{0.5cm}

 One of the most fruitful features of the Interacting Boson Model (IBM)\cite{IBM} of
nuclei is the existence of three dynamical symmetry limits. Each represents a well defined nuclear
phase, providing analytically the exact eigenstates of the system and offering a unique tool to
deeply understand the physics involved.

A quantum system has a dynamical symmetry (DS) if the hamiltonian
can be expressed as a function of the Casimir operators of a
subgroup chain. A direct consequence of this definition is that a
system exhibiting a DS is quantum integrable and analytically
solvable. The concept of quantum integrability (QI), though
sometimes associated with that of dynamical symmetry, is more
general, however. It can be stated as follows: A quantum system is
integrable if there exists a complete set of hermitian,
independent and commuting operators (constants of motion).
Clearly, if a quantum system has a DS, the Casimir operators of
its subgroups play the role of the constants of motion, fulfilling
the definition of QI. But, by no means must a system have a DS to
be quantum integrable.

Perhaps the most striking manifestation of quantum integrable
systems is the absence of level repulsion in their energy spectra.
This can be clearly seen in the IBM by looking at the transitions
between the three dynamical symmetry limits. Using the
Consistent-Q hamiltonian\cite{CW},
\begin{equation}
H=xn_{d}+\frac{x-1}{N}Q^{\chi }\cdot Q^{\chi }\label{H1}
\end{equation}
where $n_{d}=\sum_{m}d_{m}^{\dagger }d_{m}$ and $Q_{m}^{\chi
}=\left[ d_{m}^{\dagger }s+\left( -\right) ^{m}s^{\dagger
}d_{-m}\right] +\chi \left[ d^{\dagger }\widetilde{d}\right]
_{m}^{2}$, the transitions between the dynamical symmetries can be
readily explored.

In Fig. 1A, we plot the $0^{+}$ energy levels associated with the transition
from $SU(3)$ ($x=0,~\chi =-\sqrt{7}/2$%
) to $O(6)$ ($x=0,~\chi =0$) and with the transition from $O(6)$
to $\overline{SU(3})$ ($x=0,~\chi =\sqrt{7}/2$). The results are
shown as a function of $\chi $ and for $N=10$ bosons. The level
repulsion between pairs of energy levels is present everywhere,
except at the $\chi =0$ point where the $O(6)$ DS is realized. At
this point, due to the absence of level repulsion, level crossings
are allowed. In the inset we amplify the first of the two allowed
level crossings.

In Fig. 1B we show the analogous results for the $0^{+}$ states of
a system of $N=10$ bosons along the transition from $SU(3)$
($x=0,~\chi =-\sqrt{7}/2$) to $U(5)$ ($x=1$), as a function of the
parameter $x$. The level repulsion phenomenon appears any time two
levels are close enough. The inset amplifies one of the closest
two level approaches, showing that level repulsion prevents the
crossing. The level repulsion between the ground state and the
first excited $0^{+}$ state leads in the thermodynamic limit to a
non-analytic behavior of the ground state energy, defining a
critical point characterized by a first-order phase
transition\cite{Zam}.

An interesting question is what is the structure and origin of the
phase transition from $SU(3)$ to $\overline {SU(3)}$. As noted
earlier, this transition proceeds through the $O(6)$ DS. As also
noted earlier, level crossings are permitted at that {\em critical
symmetry} point. Thus, if the $SU(3) $ to $\overline {SU(3)}$
transition is indeed first-order, it is interesting to ask how
this can come about in the absence of level repulsion at the
critical point. Further study on this issue is clearly required.

\begin{figure}
\epsfysize=7cm \epsfxsize=12cm \hspace{2.5cm} \epsffile{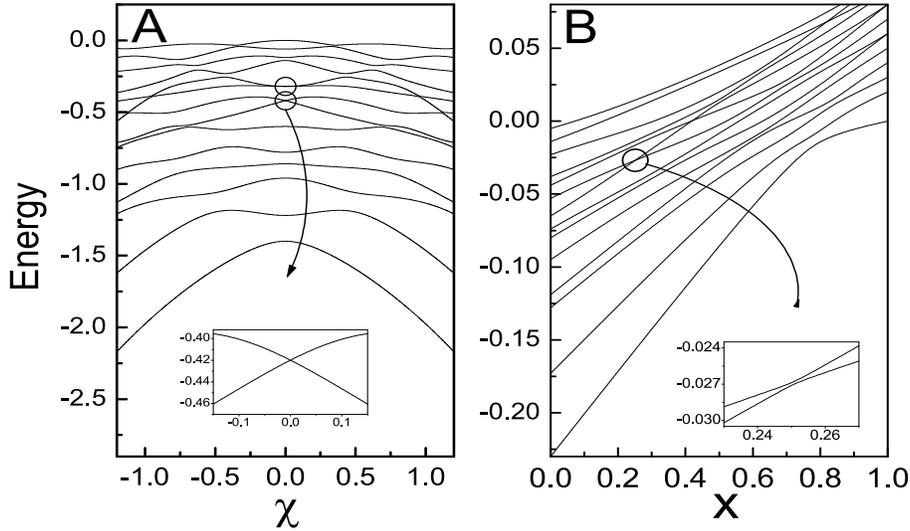}

\caption{ $0^{+}$ energy levels for a system of $N=10$ bosons as a
function of the parameter $\chi$ for the $SU(3)$-$O(6)$-$\overline
{SU(3)}$ transition (1A), and as a function of the parameter $x$
for the $SU(3)$-$U(5)$ transition (1B)} \label{fig1}
\end{figure}

The transition from $O(6)$ to $U(5)$ is described by the
hamiltonian (\ref {H1}) by setting $\chi =0$ and varying $x$ from
$0$ to $1$. Earlier investigation of the properties of the
spectrum for this leg of the Casten triangle made clear that the
system is integrable everywhere on the path, even though there is
no global DS\cite{AW,WA,CJ}. To facilitate discussion of this
region of parameter space, it is useful to digress for a moment
and to discuss a class of integrable boson pairing models. As we
will see, the $O(6)$ to $U(5)$ leg of the Casten triangle falls
within this class of models.

The models of interest are among three classes of integrable
models for fermion and boson systems\cite{DES}. Though integrable,
these models do not in general relate to a DS. The models we will
be discussing  are based on the $SU(1,1)$ pair algebra

\begin{equation}
K_{l}^{0}=\frac{1}{2}\sum_{m}\left( b_{lm}^{\dagger }b_{lm}+\frac{1}{2}%
\right) ,~K_{l}^{+}=\sum_{m}b_{lm}^{\dagger }b_{l\overline{m}}^{\dagger
}=\left( K_{l}^{-}\right) ^{\dagger }  \label{K}
\end{equation}
where the operators $b_{lm}^{\dagger }\left( b_{lm}\right) $
create (destroy) a boson in state $lm$. In terms of the generators
(\ref{K}), the constants of motion of the {\em rational} class are

\begin{equation}
R_{l}=K_{l}^{0}+2g\sum_{l^{\prime }\left( \neq l\right) }\frac{1}{\eta
_{l}-\eta _{l^{\prime }}}\left[ \frac{1}{2}\left( K_{l}^{+}K_{l^{\prime
}}^{-}+K_{l}^{-}K_{l^{\prime }}^{+}\right) -K_{l}^{0}K_{l^{\prime }}^{0}%
\right]  \label{R}
\end{equation}

It can be readily confirmed that the operators $R_{l}$ are
hermitian, are independent (none of them can be expressed as a
function of the others), and mutually commute. Moreover, if the
system has $M$ single boson states $l$, there are as many
operators (\ref{R}) as quantum degrees of freedom, constituting a
complete set of constants of motion. The pairing strength $g$ and
the set of $M$ real numbers $\eta _{l}$ are free parameters.

If a system is quantum integrable there should exist a unique
basis of common eigenstates of the $M$  operators $R_{l}$. It has
been shown\cite{DES} that this complete set of eigenvectors can be
formally written as a product of boson pairs acting on a subspace
of unpaired boson states $\left| \nu \right\rangle \equiv \left|
\nu _{1},\nu _{2},\cdots ,\nu _{M}\right\rangle $

\begin{equation}
\left| \Psi \right\rangle =\prod_{\alpha =1}^{n}\left( \sum_{l=1}^{M}\frac{1%
}{2\eta _{l}-e_{a}}K_{l}^{+}\right) \left| \nu \right\rangle  \label{Psi}
\end{equation}
where the $n$ parameters $e_{\alpha }$ that apply to each
eigenstate are particular solutions of the set of $n$ coupled
nonlinear equations

\begin{equation}
1+g\sum_{l}\frac{2l+2\nu _{l}+1}{2\eta _{l}-e_{\alpha }}-4g\sum_{\beta
\left( \neq \alpha \right) }\frac{1}{e_{\alpha }-e_{\beta }}=0  \label{Equ}
\end{equation}

The total number of bosons is $N=2 n+\sum_l \nu_l$ and the
eigenvalues $r_{l}$ of the $R_{l}$ can be found in Ref. [7].


Any hamiltonian constructed as a linear combination of the
constants of motion (\ref{R}), viz: $H=2\sum_{l}\epsilon
_{l}R_{l}\left( g,\eta \right)$,
commutes with them, and thus is diagonal in the basis of
eigenstates (\ref{Psi}). The hamiltonian eigenvalues are the same
linear combination $2\sum_{l}\epsilon _{l}r_{l}\left( g,\eta
\right)$. In particular, the boson pairing hamiltonian that was
solved by Richardson\cite{Richa} is obtained by choosing the
coefficients of the linear combination equal to the parameters
inside the $R_l$ operators, $\epsilon_l=\eta_l$. We will use the
following parameterization of the pairing hamiltonian:

\begin{equation}
H_{P}=\frac{x}{2}\sum_{lm} l b_{lm}^{\dagger }b_{lm}+\frac{1-x}{2
N }\sum_{lml^{\prime
}m^{\prime }}b_{lm}^{\dagger }b_{l\overline{m}}^{\dagger }b_{l^{\prime }%
\overline{m}^{\prime }}b_{l^{\prime }m^{\prime }}  \label{HP}
\end{equation}
for which the eigenvalues are

\begin{equation}
E_{P}=\sum_{l}\eta _{l}\nu _{l}+\sum_{\alpha =1}^{M}e_{\alpha }  \label{ep}
\end{equation}
\

With the limitation to $s$ and $d$ bosons, the hamiltonian
(\ref{HP}) describes a transition from a spherical vibrational
$U(5)$  DS to the $\gamma $ unstable $O(6)$ DS. In Fig. 2 we show
the corresponding $0^{+}$ states for $N=10$ bosons as a function
of $x$. In the limit of $x=0$, we are in the exact $O(6)$ DS
limit; for $x=1$ we are in the $U(5)$ DS limit. As previously
discussed, the behavior of the levels as a function of $x$ does
not show any sign of level
repulsion. Since the $O(5)$ Casimir operator commutes with the hamiltonian (%
\ref{HP}), the eigenstates can be labelled by the boson seniority
quantum number $\tau $.

In the left panel, we plot all the $0^+$ levels, with the
seniority quantum number $\tau$ specified on the right vertical
axis. There are several level crossings in the figure, but they
all correspond to pairs of levels with different seniority quantum
numbers. The $0^+$ levels with $\tau=0$ are displayed in the right
panel. These levels evolve independently with the parameter $x$.
The fact that the complete transitional region from $O(6)$ to
$U(5)$ is quantum integrable has been been previously
noted\cite{CJ}. From the six quantum degrees of freedom of the
$U(6)$ dynamical group, four of them are taken into account by the
Casimir operators of the $O(5)$ subgroup chain, the fifth is the
conserved number of bosons $N$ and the sixth is any linear
combination of the Casimir operators of $O(6)$ and $U(5)$, e.g.
the hamiltonian, which by construction commutes with all the other
constants of motion. We arrive to the same conclusion from the
constants of motion given in (\ref{R}). In an $sd$ space there are
two constants of motion, $R_{s}$ and $R_{d}$. The sum gives the
boson number $N$ and any other combination defines a hamiltonian
interpolating between $O(6)$ and $U(5).$ The states with unpaired
bosons are completely classified by the $O(5)$ subgroup.

\begin{figure}

\epsfysize=8cm \epsfxsize=12cm \hspace{2.5cm}\epsffile{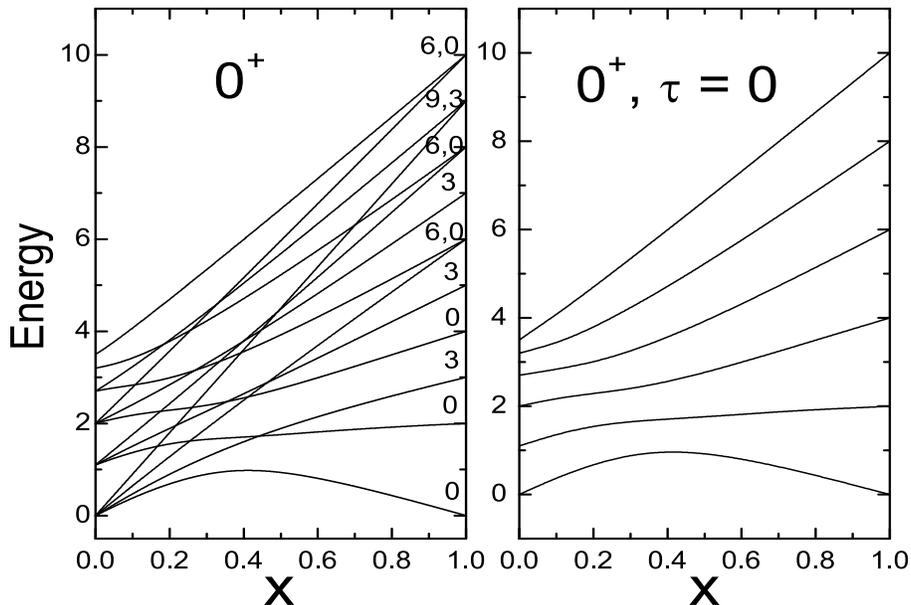}


\caption{ $0^{+}$ energy levels for a system of $N=10$ bosons as a
function of the parameter $x$ for the $O(3)$-$U(5)$ transition.
The left panel shows all $0^{+}$ levels; the right panel only the
$\tau = 0$ levels.} \label{fig2}
\end{figure}
A non-analytic point in the thermodynamic limit due to level
repulsion is precluded in the transition from $O(6)$ to $U(5)$.
The only source of non-analyticity within an integrable region is
a level crossing. However, as we see in Fig. 2, states with $\tau
=0$ do not cross. The exact solvability of the model implies that
the energy is an analytic function of the parameter $x$ and only
phase transitions of order greater than one are permitted. We may
wonder, therefore, under what conditions are crossings between
$\tau =0$ states possible. It can be shown that crossings between
states with the same set of quantum numbers can take place when
there are at least two parameter dependent constants of motion.
The simplest example is a boson model with $sdg$ bosons. In this
case, there are three independent constants of motion, $R_s$,
$R_d$ and $R_g$, but factorizing out the boson number $N$ as the
sum of the three, we are left with two that are parameter
dependent. An analysis based on the structure of the $U(15)$
dynamical group of the $sdg$ IBM leads to the same conclusion. The
pairing hamiltonian (\ref{HP}) is a linear combination of the
Casimir operators of two subgroups, those associated with $U(14)$
and $O(15)$. Both subgroups constitute a DS of the $U(15)$
dynamical group and both have the group $O(14)$ as a common
subgroup. In Fig. 3 we show the $0^{+},~\tau =0$ states for $N=10$
bosons as a function of the parameter $x$ for this model. As
expected, there are no signs of level repulsion, but there are
level crossings. However, there are no crossings with the ground
state, which would have been evidence for a first-order phase
transition, even though it is a finite system.

\begin{figure}

\epsfysize=8cm \epsfxsize=12cm \hspace{2.5cm} \epsffile{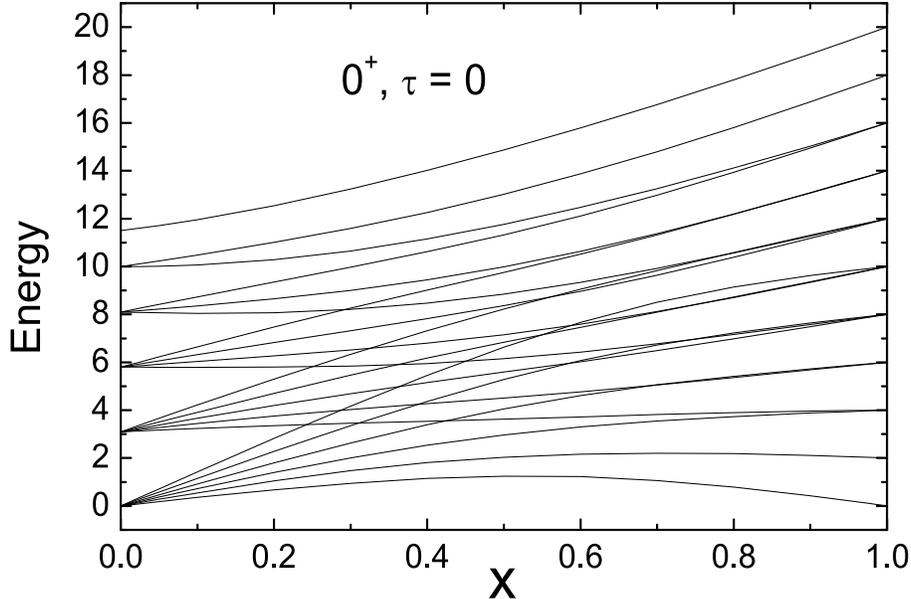}


\caption{ $sdg$ $0^{+}, \tau=0$ energy levels for a system of $N=10$ bosons as a function of the
parameter $x$ for the $O(15)$-$U(14)$ transition.} \label{fig3}
\end{figure}

In previous works\cite{PR1,PR2}, we showed that the rational class
of integral models with repulsive pairing exhibits a quantum phase
transition (QPT) to a symmetry broken phase with a macroscopic
occupation of the two lowest single-boson states. This led us to
conclude that repulsive pairing between bosons is a new and robust
mechanism for enhancing $sd$ dominance\cite{PR2} in interacting
boson models of nuclei.

Within the Landau theory, a second-order phase transition is
related to a continuous change of the order parameter from $0$ in
the disordered phase to a non-zero value in the ordered phase. The
order parameter is not unique, as its definition depends on the
particular problem. In the Ehrenfest approach, the order of the
transition is related to the order of the first discontinuous
derivative of the energy.

\begin{figure}
\epsfysize=8cm \epsfxsize=6cm \hspace{6cm}
\epsffile{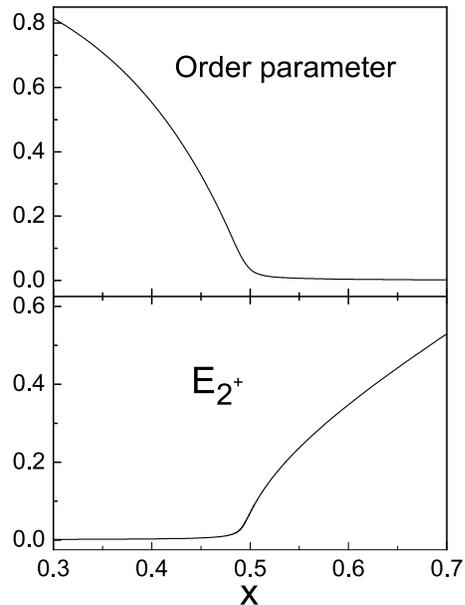}
\caption{ Order parameter (upper panel) and $2^+$ energy (lower panel) as a function of $x$ for a
system of $N=1000$ bosons in the $O(6)$-$U(5)$ transition.} \label{fig4}
\end{figure}

The advantage of an exactly solvable model is that we can reconcile these two approaches to phase
transitions, by defining the order parameter as the derivative of the hamiltonian (\ref{HP}) with
respect to $x$. Making use of the Helmann-Feyman theorem, the order parameter is given by
\[
O=\frac{\partial H}{\partial x}=n_{d}-\frac{1}{N}\left( d^{\dagger }\cdot
d^{\dagger }-s^{\dagger }s^{\dagger }\right) \left( d\cdot d-ss\right)
~\rightarrow ~\left\langle O\right\rangle =\frac{\partial E}{\partial x}
\]

In Fig. 4 we show the order parameter (we have subtracted the
constant $N-1$ to assure it is zero in the disordered phase)  and
the energy of the first excited state with angular momentum
$2^{+}$, for a system of $N=1000$ bosons and in the vicinity of
the phase transition. From fig. 4, we see that the phase
transition is of second order. By including finite size
corrections, we can readily study the detailed properties of this
phase transition, as well as the corresponding phase transitions
in larger spaces. The present analysis based on the exact solution
for very large systems confirms recent results\cite{Jo} based on
the phenomenological Landau theory on the characteristics of this
phase transition.


\begin{thebibliography}{99}

\bibitem{IBM}F. Iachello and A. Arima, {\em The Interacting Boson
Model of Nuclei} (Cambridge University Press, Cambridge, UK.
1987).

\bibitem{CW}R. F. Casten and D. D. Warner, {\em Rev. Mod. Phys.} {\bf 60}, 389 (1988).

\bibitem{Zam} N.V. Zamfir {em et. al.}, {\em Phys. Rev.} C {\bf 66},
021304 (2002).

\bibitem{AW} Y. Alhassid and N. Whelan, {\em Phys. Rev. Lett.} {\bf
67}, 816 (1991).

\bibitem{WA} N. Whelan and Y. Alhassid, {\em Nucl. Phys.} A {\bf 556}, 42 (1993).

\bibitem{CJ} P. Cejnar and J. Jolie, {\em Phys. Rev.} E {\bf 58},
387 (1998).

\bibitem{DES} J. Dukelsky, C. Esebbag and P. Schuck, {\em Phys.
Rev. Lett.} {\bf 87}, 066403 (2001).

\bibitem{Richa} R. W. Richardson, {\em J. Math. Phys.} {\bf 9},
1327 (1968).

\bibitem{PR1} J. Dukelsky and P. Schuck, {\em Phys. Rev. Lett.} {\bf 86}, 4207 (2001).

\bibitem{PR2} J. Dukelsky and S. Pittel, {\em Phys. Rev. Lett.} {\bf 86}, 4791 (2001).

\bibitem{Jo} J. Jolie {\em et. al.}, {\em Phys. Rev. Lett.} {\bf 86}, 4791 (2001).


\end{thebibliography}
\end{document}